\pdfoutput=1
\documentclass[sigconf, anonymous]{article}
\usepackage{ijcai24}

\usepackage{times}
\usepackage{soul}
\usepackage{url}
\usepackage[hidelinks]{hyperref}
\usepackage[utf8]{inputenc}
\usepackage[small]{caption}
\usepackage{graphicx}
\usepackage{amsmath}
\usepackage{amsthm}
\usepackage{booktabs}
\usepackage{algorithm}
\usepackage{algorithmic}
\usepackage[switch]{lineno}

\usepackage{tabularx}
\usepackage{upgreek}
\usepackage{array}
\usepackage{multirow}
\usepackage{amssymb}
\usepackage{comment}
\usepackage{makecell}
\usepackage{subfigure}

\title{Global Tropical Cyclone Intensity Forecasting with Multi-modal Multi-scale Causal Autoregressive Model}

\author{
    Xinyu Wang$^{1,2}$\and
    Kang Chen$^{1,2}$\and
    Lei Liu$^{1}$\thanks{Corresponding Author.}\and
    Tao Han$^2$\and
    Bin Li$^1$\and
    Lei Bai$^{2\ast}$ \\
    \affiliations
    $^1$University of Science and Technology of China\\
    $^2$Shanghai Artificial Intelligence Laboratory\\
    \emails
    \{xinyuwang, ck6\}@mail.ustc.edu.cn,
    liulei13@ustc.edu.cn,
    hantao10200@gmail.com,
    binli@ustc.edu.cn,
    bailei@pjlab.org.cn,
}

\begin{document}
\pagestyle{plain}
\maketitle
\begin{abstract}
Accurate forecasting of Tropical cyclone (TC) intensity is crucial for formulating disaster risk reduction strategies. Current methods predominantly rely on limited spatiotemporal information from ERA5 data and neglect the causal relationships between these physical variables, failing to fully capture the spatial and temporal patterns required for intensity forecasting. To address this issue, we propose a Multi-modal multi-Scale Causal AutoRegressive model (MSCAR), which is the first model that combines causal relationships with large-scale multi-modal data for global TC intensity autoregressive forecasting. Furthermore, given the current absence of a TC dataset that offers a wide range of spatial variables, we present the Satellite and ERA5-based Tropical Cyclone Dataset (SETCD), which stands as the longest and most comprehensive global dataset related to TCs. Experiments on the dataset show that MSCAR outperforms the state-of-the-art methods, achieving maximum reductions in global and regional forecast errors of 9.52\% and 6.74\%, respectively. The code and dataset are publicly available at~\href{https://anonymous.4open.science/r/MSCAR}{https://anonymous.4open.science/r/MSCAR}.
\end{abstract}
\begin{table*}
    \resizebox{\textwidth}{!}{
    \centering
    \begin{tabularx}{\linewidth}{lXXXX} 
        \toprule
                              &  SETCD           & Digital Typhoon & HURSAT   & TCIR \\
        \midrule
        Temporal coverage     & 1980-2022        & 1978-2022       & 1978-2015 & 2003-2017    \\
        Temporal resolution   & three hours        & one hour        & three hours  & three hours \\
        Spatial coverage      & Global coverage  & Western North Pacific   & Global coverage & Global coverage \\
        Spatial resolution    & \makecell[tl]{$\sim$8km (satellite) \\ $\sim$30km (ERA5)}
        & $\sim$5km   & $\sim$8km   & $\sim$8km     \\
        \makecell[tl]{Image coverage \\ (pixels)}       & \makecell[tl]{572$\times$572 (satellite) \\ 160$\times$160 (ERA5)}  
        & 512$\times$512  & 301$\times$301   & 201$\times$201     \\
        Variables     & infrared, water vapor, visible,~\textbf{ERA5 (69 variables)}
        & infrared (others on the Website) & infrared, water vapor, visible, near IR, split window    & infrared, water vapor, visible, passive microwave    \\
        TCs                   & \textbf{4,668}               & 1,099             & 3,946   & 1,285   \\
        Frames                & \textbf{282,505}            & 189,364            & 237,516    & 70,501   \\
        \bottomrule
    \end{tabularx}}
    \caption{Comparison of SETCD with existing TC-related datasets.}
    \label{tab:datasets_cmp}
\end{table*}
\section{Introduction}
Tropical cyclones (TCs) rank among the most destructive natural disasters. They pose substantial risks to maritime navigation and are frequently accompanied by disastrous precipitation, triggering events such as floods and landslides~\cite{tc_disaster}. Consequently, the timely and accurate forecasting of cyclone intensity holds paramount importance in mitigating economic losses and minimizing human casualties.
%Despite research on the formation and development of TCs dating back to the last century~\cite{tc_genesis}, the dynamics of TCs are complex, influenced by various environmental factors, and represent a highly nonlinear chaotic system~\cite{tc_env_factors}. 
Despite research on the formation and development of TCs dating back to the last century~\cite{tc_genesis}, the understanding of TCs' physical processes remains shrouded in numerous enigmas~\cite{tc_env_factors}. How to achieve higher accuracy in TC intensity forecasting has remained a challenging topic. 

TC intensity determination relies on key parameters like maximum sustained wind (MSW) and minimum sea level pressure (MSLP). In the early stages, Dvorak proposed the Dvorak technique~\cite{dvorak}, which utilizes satellite images to estimate and forecast TC intensity, and its variants are still widely used by forecasters worldwide. Despite its widespread use, the technique's subjective nature leads to unsatisfactory results. Now many forecast agencies have developed regional or global TC intensity forecast models based on statistical techniques and simplified dynamical models, commonly referred to as numerical weather prediction (NWP)~\cite{nwp}. However, due to the complex dynamics of TCs, these NWP models often require significant computational resources and time~\cite{ruttgers2019prediction}. In recent years, deep learning methods have made significant advancements in the field of TC research~\cite{TC_estimation,TC_zili,pangu}. Compared to traditional NWP models, deep learning networks have demonstrated their suitability for learning complex nonlinear systems with reduced computational requirements. Researchers have explored deep learning methods for TC intensity forecasting, which can be broadly categorized into three main approaches: 

\begin{enumerate}
    \item A straightforward approach, as demonstrated in studies~\cite{mmstn,transformer_track_intensity_forecasting}, is to consider it as a time series forecasting problem.
    \item By considering it as a spatial-temporal sequence forecasting task and incorporating atmospheric variables, researchers have introduced deep learning networks to learn underlying physical relationships and achieve more accurate forecasts~\cite{TC_Pred,tc_probabilistic,mgtcf}.
    \item In recent studies~\cite{cmp_pangu}, it has been regarded as a downstream task, where researchers approach it by forecasting atmospheric variable fields and subsequently applying post-processing techniques to derive intensity forecasts.
\end{enumerate}
While the aforementioned methods have achieved a certain degree of success, they still face challenges in terms of accuracy, flexibility, and real-time performance. The first approach overlooks the inherent physical characteristics of TCs, as they are highly coupled with various atmospheric variables. The life cycles of TCs vary in length and are non-fixed, meaning the input length is often constrained in practical applications, allowing for the possibility of identical input intensities but completely different subsequent intensities. The second approach primarily relies on a limited amount of the fifth generation of European Centre for Medium-Range Weather Forecasts(ECMWF) reanalysis data (ERA5) to provide spatial information and overlook the causal relationships that exist in the temporal dimension of these data. This limitation hampers the modeling of spatiotemporal correlations and hinders the achievement of high-precision TC intensity forecasting. The third approach has great potential but it requires enormous data and computational resources for training, and since intensity is not directly predicted, forecast errors could be larger~\cite{cmp_pangu}. Additionally, there are cases where the predicted variables do not align perfectly with the desired outputs~\cite{pangu,graphcast,fengwu,fuxi,cmp_pangu}. 
%\footnote{In ERA5, the available data is for ``10m wind'' rather than MSW.}

Influenced by~\cite{pangu} and the traditional Dvorak technique~\cite{dvorak}, we consider using satellite imagery and ERA5 data~\cite{era5} as spatial input to deeply analyze historical TC intensity information and infer future intensities. Building upon these principles, we propose the Multi-modal multi-Scale Causal AutoRegressive model (MSCAR). MSCAR employs a multi-scale information fusion technique, utilizing a Feature Pyramid Network (FPN)~\cite{fpn} to combine satellite imagery data and ERA5 data. The spatial information is segmented into patches similar to Vision Transformer (ViT)~\cite{vit}. By incorporating causal cross-attention with historical TC intensity sequences, MSCAR adeptly captures the intricate interplay between spatial and temporal dynamics. To the best of our knowledge, this is the first global TC intensity forecasting model that employs an autoregressive approach based on satellite imagery and ERA5 data. Our work offers noteworthy contributions in the following aspects:
\begin{itemize}
    \item We introduce MSCAR, a novel network that incorporates causal relationships for global TC intensity autoregressive forecasting on a large-scale multimodal dataset.
    \item We provide the SETCD, the longest and most comprehensive global TC dataset, comprising 43 years of data, 4,668 TCs, and 72 variables in total.
    \item We conduct experiments on both a global scale and regional basins, achieving state-of-the-art results.
    \item Real-time experiments with MSCAR demonstrate its practicality and robustness, as it maintains comparable performance to non-real-time scenarios.
\end{itemize}

\begin{comment}
    \begin{figure*}[htp]
        \centering
        \includegraphics[width=0.9\textwidth]{my paper/all_tc_small.png}
        \caption{The global TC colored by their Saffir-Simpson Hurricane Wind Scale (SSHWS) intensity categories from 1980 to 2022.}
        \label{fig:all_tc}
    \end{figure*}
\end{comment}

\section{Related Work}
\paragraph{TC Intensity Forecast} In regional basins, there has been an abundance of research utilizing deep learning methods for TC intensity forecast. By leveraging the differential results of historical TC information, ~\cite{transformer_track_intensity_forecasting} explores the correlation within time series to enhance the performance of 24-hour short-term forecasts. ~\cite{TC_Pred} proposed a TC intensity prediction framework called TC-Pred. This framework employs an encoder-decoder structure to extract historical state information effectively. It leverages attention mechanisms to enhance the interaction between the historical state information in the encoder and the future sequential state information predicted by the decoder. In contrast to deterministic forecasting approaches, ~\cite{tc_probabilistic} delves into uncertainty in TCs using deep learning methods. They introduce a novel approach that utilizes multimodal data to predict the mean and distribution of intensity, enabling probabilistic forecasts of TC intensity for 24-hour short-term forecasts. ~\cite{mmstn} adeptly employs a GAN network structure to generate multiple TC outcomes, essentially achieving ensemble forecasting. Building upon this foundation, they further advance their approach and propose the MGTCF~\cite{mgtcf}, a highly efficient utilization of heterogeneous meteorological data. Following in the footsteps of Social GAN~\cite{Social_gan}, the approach achieves outstanding results by selecting the output with minimal error compared to the ground truth values from multiple generated results as the ultimate output. 

On a global scale, ~\cite{pangu,graphcast,fengwu,fuxi} have leveraged ERA5 data to train models that predict future atmospheric variables for up to one week or even longer, based on atmospheric variables from the previous time step. These models have exhibited exceptional performance, surpassing ECMWF in nearly all aspects. Subsequently, referencing ~\cite{cmp_pangu}, the predicted fields and ECMWF's tracking algorithm can be utilized to determine the intensity information of TC accurately.

\paragraph{TC dataset} 
\begin{comment} 
A dataset containing all information collected throughout the entire lifecycle of TCs, analyzed to obtain the best estimate, is commonly referred to as the best track dataset.~\cite{IBTrACS}. Renowned institutions such as the China Meteorological Administration's Shanghai Typhoon Institute (CMA), the Japan Meteorological Agency (JMA), and NOAA's National Hurricane Center (NHC) each possess their own best track datasets. It is important to note that due to the scarcity of ground-based observations for TCs and variations in techniques used for estimating their positions and intensities, there are slight disparities even in the definition of cyclone intensity among these organizations. As a result, variations in the best track datasets for TCs exist to some extent, reflecting the unique approaches and methodologies employed by each institution~\cite{chen2019verification}. On a global scale, the International Best Track Archive for Climate Stewardship (IBTrACS) amalgamates estimates from multiple institutions, offering a comprehensive and extensive repository of global TC information. While observational data may not be uniform across all sources, IBTrACS is a relatively reasonable and widely accepted best estimate~\cite{IBTrACS}. In our experiments, we consider the IBTrACS dataset as the ground truth, given its reliability and availability for assessing TC characteristics.
\end{comment}

For TC image datasets, the customary approach is to use the TC center as the image center and select a bounding box encompassing a sufficient surrounding area. This allows observation of TCs and environmental variables at varying scales within the chosen range. 

On a global scale, one of the most widely recognized and influential datasets is the Hurricane Satellite (HURSAT) dataset provided by NOAA~\cite{HURSAT}, featuring 8km resolution satellite imagery from 1978 to 2015 at 3-hour intervals. Its extensive coverage and high-quality imagery make it indispensable for studying and understanding these impactful weather systems. TCIR is a dataset introduced in~\cite{TCIR} that offers distinct channels separate from HURSAT but with a relatively short time coverage. For the Western North Pacific basin, an esteemed contribution is Digital Typhoon~\cite{Digital_Typhoon}with longer duration and finer spatiotemporal resolution in satellite imagery. This exceptional dataset allows for a comprehensive exploration of TCs in the Western North Pacific, unveiling their evolution and characteristics over an extended period. 

In Table~\ref{tab:datasets_cmp}, we present a comparison between our SETCD and the aforementioned datasets. This detailed analysis allows for a comprehensive understanding of each dataset's strengths and distinctive characteristics, enabling researchers and practitioners to make informed decisions based on their specific requirements. By providing this overview, we aim to facilitate a deeper appreciation of diverse TC datasets and their respective contributions.

\begin{figure*}[htp]
    \centering
    \includegraphics[width=0.84\textwidth]{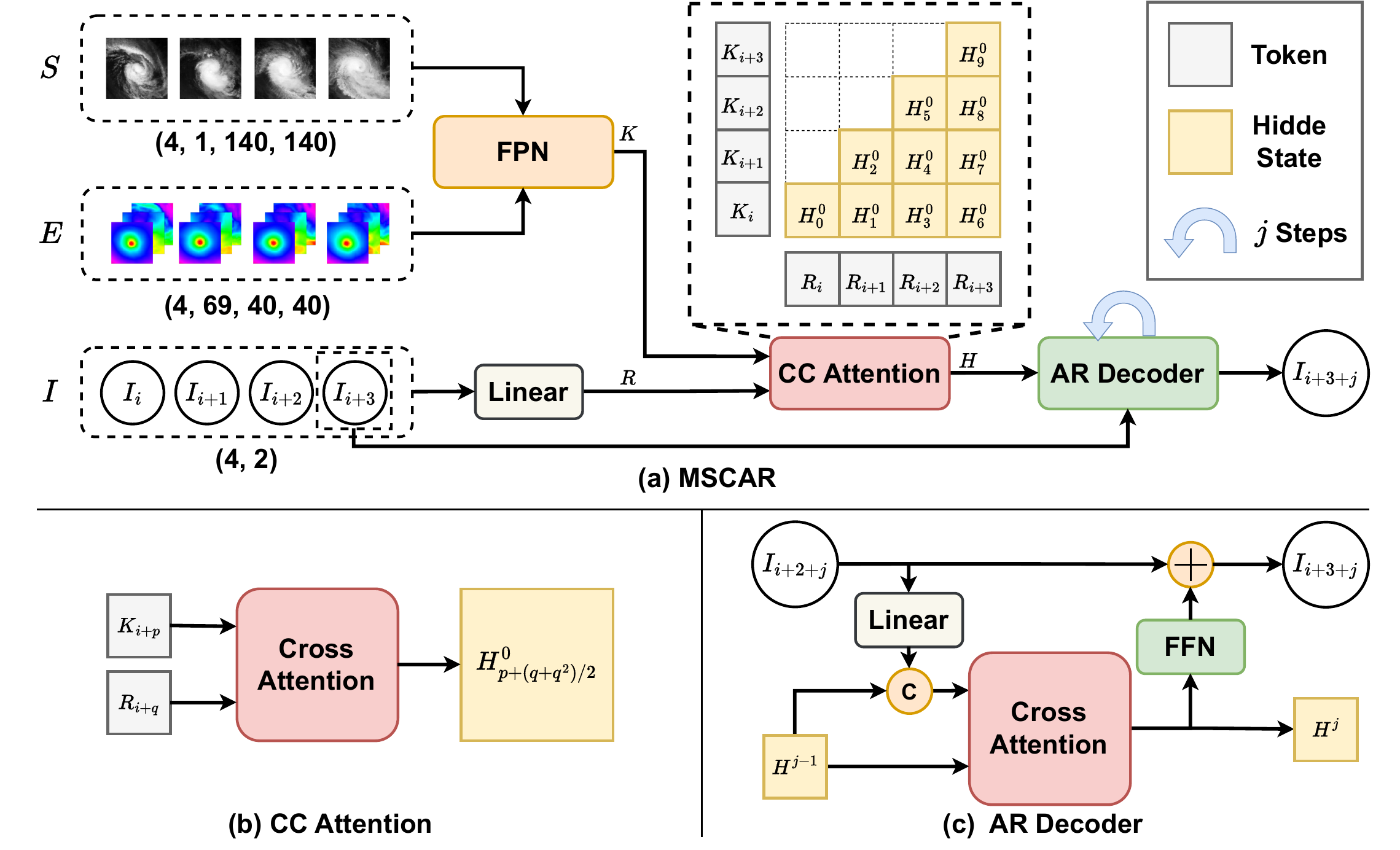}
    \caption{(a) The overall architecture of the MSCAR. The diagram illustrates the process of forecasting the TC intensity at the $j$th time step in the future based on the previous four-time steps. ${K}$ represents the tokens for spatial information, and ${R}$ represents the tokens for temporal information. These tokens are pairwise matched using CC Attention, shown by the dashed box. This process generates ${H}$, which is then passed through the AR Decoder for ${j}$ iterations. Finally, the AR Decoder produces the TC intensity for the future ${j}$ step. (b) The diagram illustrates the calculation method of CC Attention (c) The AR Decoder takes the initial ${H}$ and iteratively updates it to obtain the TC intensity.}
    \label{fig:MSCAR}
\end{figure*}

\section{Methodology}

\subsection{Problem Definition}

MSCAR aims to forecast future ${M}$ steps of TC intensities based on historical ${N}$ steps of satellite data, ERA5 data, and TC intensities. Satellite data at time ${i}$ are represented as ${S_i}$, while ERA5 data and TC intensities are denoted by ${E_i}$ and ${I_i}$ respectively. The main goal of MSCAR is to design and implement a function $f$ that can be expressed as follows:
\begin{align}
    \displaystyle
    {\hat{I}_{i+N:i+N+M-1}} = f(~{S_{i:i+N-1}}, ~{E_{i:i+N-1}}, ~{I_{i:i+N-1}}, ~{M})
\end{align}%

where ${\hat{I}_{i+N:i+N+M-1}}$$=\{{\hat{I}_{i+N}}, $${\hat{I}_{i+N+1}}$$,\cdots,$${\hat{I}_{i+N+M-1}}\}$ denotes the forecast of the future ${M}$ steps of TC intensity. In this paper, we set the value of ${N}$ to 4, while M varies depending on the specific requirements.

\begin{comment}
    \paragraph{GridSat-B1} The Geostationary IR Channel Brightness Temperature (BT)- GridSat-B1 Climate Data Record (CDR) ~\cite{gridsat1,gridsat2}, with a temporal resolution of 3 hours and a spatial resolution of 0.07$^\circ$$\times$0.07$^\circ$, encompasses global (70N to 70S) infrared window brightness temperatures from geostationary Infrared (IR) satellites, spanning from 1980 to the present day. It comprises the following three channels:
    \begin{itemize}
        \item CDR-quality infrared window channel (near 11 $\upmu$m).
        \item Infrared water vapor channel (near 6.7 $\upmu$m).
        \item Visible channel (near 0.6 $\upmu$m).
    \end{itemize}
    
    \paragraph{ERA5} The fifth-generation ECMWF atmospheric reanalysis, represents a comprehensive global climate dataset spanning from January 1940 to the present. With a spatial resolution of 30km, ERA5 offers hourly estimates of numerous atmospheric, land, and oceanic climate variables. It utilizes 137 vertical levels, covering the Earth's atmosphere from the surface up to 80km in height, ensuring a high-resolution representation of atmospheric conditions. 
    
\end{comment}

\subsection{SETCD}
We have curated a vast collection of spatial information data about TC from 1980 to 2022, which we have named the Satellite and ERA5-based Tropical Cyclone Dataset (SETCD). Our dataset is derived from GridSat-B1~\cite{gridsat1,gridsat2} and ERA5~\cite{era5}. In addition to the above data sources, we require best track datasets providing TC latitude, longitude, intensity, and other data. Globally, the International Best Track Archive for Climate Stewardship (IBTrACS)~\cite{IBTrACS} integrates estimates from multiple agencies to provide more extensive global TC information. Therefore, we uniformly use IBTrACS as the best track dataset.

To capture TC-related information, SETCD uses latitude and longitude positions from IBTrACS as center points. We extract a $40^\circ$ diameter region from GridSat-B1 and ERA5 datasets at a three-hour temporal resolution to ensure comprehensive coverage. Our objective is to include all documented TCs for future research, even if unnamed or low-pressure systems in IBTrACS are incorporated. This inclusive approach ensures that all TCs are considered, enabling thorough exploration in the future without overlooking any cases.

%Since the majority of the IBTrACS dataset provides optimal estimates of TC positions at 6-hour intervals~\cite{IBTrACS}, we sample data at 00:00, 06:00, 12:00, and 18:00 UTC for each day to balance storage efficiency and representation accuracy.
%Moreover, to address the potential issue of duplicate occurrences of the same TC name, SETCD utilizes a unique storm identifier (SID) assigned by the IBTrACS algorithm as the naming convention for storage. 

In summary, the SETCD dataset comprises 4,668 TCs from 1980 to 2022, consisting of 282,505 satellite and ERA5 instances extracted around TC centers. The satellite data within the SETCD dataset consists of three channels from GridSat-B1: infrared, water vapor, and visible. 
% The infrared and water vapor channels have a completeness rate of over 90\% %, indicating high data availability. However, the visible channel is approximately 35\% complete, suggesting some missing values. To address these incomplete portions, all missing values have been uniformly treated and replaced with NaN (not a number) values during data processing. 
The ERA5 data in the SETCD dataset is selected based on the same criteria as Pangu-Weather~\cite{pangu}, including 69 factors. Table~\ref{tab:datasets_cmp} compares SETCD to other TC-related datasets, highlighting key aspects. Additional dataset details are provided in the Appendix. 

\begin{comment}
    Figure~\ref{fig:all_tc} presents all TCs and their intensity categories involved in this study. The intensities are classified according to the Saffir-Simpson Hurricane Wind Scale (SSHWS), and the data is sourced from IBTrACS.
\end{comment}

Additionally, considering that ERA5 data is not available in real-time, in practical applications, analysis data can be considered as a suitable substitute for ERA5 data, as it can be approximated in real-time. Therefore, we have also made available analysis data from 2010 to 2022 for studying real-time performance.

\subsection{MSCAR}
 
The proposed MSCAR model is a multimodal framework that makes use of multiple data sources to extract multiscale information and employs autoregressive modeling in the latent space. The framework is illustrated in Figure~\ref{fig:MSCAR}, showcasing the architecture and workflow of the model. Our approach involves utilizing a FPN to merge ERA5 and satellite data, enabling the extraction of multi-scale features that capture TC spatial information. Shallow temporal information is extracted from the TC intensity sequence using linear layers. This extracted spatiotemporal information is then fed into a Causal Cross-Attention Module (CC Attention), which further integrates spatial and temporal information to derive potential latent states reflecting future intensity changes. Finally, an Autoregressive Decoder module (AR Decoder) iteratively predicts TC intensity by considering the evolving states.

\paragraph{Feature Pyramid Network (FPN)} Nascent TCs are generally small-scale cyclone systems. As a TC's intensity continuously increases, its scale gradually expands. Meanwhile, surrounding atmospheric variables like specific humidity and horizontal winds also experience complex scale variations. Inspired by these observed scale evolution processes over a TC's life cycle, we employ a FPN to capture multi-scale change features. The FPN extracts characteristics reflective of influence on future intensity by representing the scale transformation resulting from a TC's strengthening over its life cycle. 

Since TCs generally have diameters within 1000km, we practically extract satellite data and ERA5 data within a $10^\circ$ range of the TC center for our experiments~\cite{TC_Pred}. For satellite data of size 140$\times$140 pixels at time ${n}$, we apply residual blocks to obtain feature maps of sizes 140$\times$140, 70$\times$70, and 35$\times$35 pixels. The ERA5 data at 40$\times$40 pixels undergoes a single residual block to obtain a 35$\times$35 pixels feature map. This feature map is then concatenated and fused with the 35$\times$35 feature map from the satellite data. Following the FPN method, we upsample the fused feature map from top to bottom to generate semantically enriched feature maps of sizes {140$\times$140, 70$\times$70, 35$\times$35, and 35$\times$35}. Inspired by ViT, we segment the feature map into patches of size 5$\times$5. These patches are then linearly embedded and concatenated with a learnable token. For each time step ${n}$, we obtain 1079 tokens, resulting in a final representation denoted as ${K\in \mathbb{R}^{N \times 1079\times C}}$, where $C$ is the dimension of the token. The representation can be expressed as Equation ~\ref{eq:FPN}. More FPN details are provided in the Appendix.
\begin{align}
    {K_n} =~{FPN}(~{S_n}, ~{E_n}) 
    \label{eq:FPN}
\end{align}%
To capture the intensity information of historical TCs, a linear layer is utilized to map the intensities of the historical TCs, resulting in ${R\in \mathbb{R}^{N \times 1\times C}}$.

\begin{comment}
    \begin{figure*}[htp]
    \centering
    \includegraphics[width=\textwidth]{my paper/CCAR.png}
    \caption{(a) CC Attention is essentially a Cross Attention module, and the diagram illustrates its computational process. (b) The iteration of ${H}$ is also achieved through a Cross Attention module, as shown in the diagram. The AR Decoder takes the initial ${H}$ and iteratively updates it to obtain the TC intensity. (c) The specific implementation of the Cross Attention module}
    \label{fig:CCAR}
\end{figure*}
\end{comment}

\paragraph{Causal Cross-Attention (CC Attention)} Having obtained the mappings of spatial information to tokens ${K}$ and the mappings of TC intensity time series to tokens ${R}$, we recognize that the representation of ${R}$ is relatively shallow and insufficient for future forecasting. To overcome this limitation, a natural and intuitive approach is to perform cross-attention computation between the semantically rich tokens ${K}$ and ${R}$ to enhance the exploration of future TC intensity information. However, TCs are inherently physical chaotic systems with temporal causal relationships. Future TC intensity is only influenced by past relevant factors. Directly using future information to calculate attention weights along with past information not only risks overfitting but may also introduce redundant noise and disrupt the physical mechanisms of TCs. Therefore, we propose Causal Cross-Attention (CC Attention), which provides a priori temporal causal relationship to reduce redundant computation while retaining the basic physical laws governing TC dynamics.

As shown in Figure~\ref{fig:MSCAR} (b), the core of CC Attention is a Cross-Attention module. Within the module, spatial information tokens ${K_{i+p}}$ at time step ${i+p}$ are used to guide the TC historical intensity sequence token ${R_{i+q}}$ at time step ${i+q}$, where ${p \leq q}$. During computation, cross-attention is mutually calculated between ${K_{i+p}}$ and $R_{i+q}$ through multiple iterations. This serves to enhance interaction and explore potential relationships. In the final iteration, the information from ${R_{i+q}}$ is used as the query, while the information from ${K_{i+p}}$ serves as the key and the value. This ultimately yields the hidden state ${H^0}$. The overall process can be formulated as:
\begin{align}
    {H_{{p+(q+q^2)/2}}^0} =~{CC}~{Atten}(~{R_{i+q}}, ~{K_{i+p}}), ~{p \leq q}
    \label{eq:CCAtten}
\end{align}
where $p,q\in [0, 1,\cdots ,N-1 ]$, $H^{0}\in \mathbb{R}^{\frac{(N+1)N}{2}\times C}$.

The Causal Cross-Attention (CC Attention) mechanism enables the model to effectively capture the spatiotemporal correlations between ${K}$ and ${R}$ by providing a priori causal relationships. Refinement and exploration are achieved through iterative computations of cross-attention, thereby obtaining more comprehensive and accurate representations.

\paragraph{Autoregressive Decoder (AR Decoder)} We aspire to enhance the flexibility and scalability of MSCAR beyond a simple fixed-output-length multiple input multiple output (MIMO) structure. Due to the discrepancy between input and output lengths, directly employing the output results for autoregression is not feasible. Taking inspiration from LSTM~\cite{LSTM}, we can introduce autoregression in the latent space, leading us to propose the AR Decoder, as illustrated in Figure~\ref{fig:MSCAR} (c).

The essence of the AR decoder is to determine whether the forecast of future TC intensity relies more on the intensity information from the previous time step or the historical spatiotemporal features. We have obtained the initial hidden state ${H^0}$, which contains information about TC intensity at the next time step. Taking the input TC intensity sequence in Figure~\ref{fig:MSCAR} (a) as an example, we input the TC intensity information, ${I_{i+3}}$, from the last time step into the AR Decoder via a linear layer to obtain a shallow representation similar to ${R}$. We concatenate it with the initial hidden state ${H^0}$ as the key and value, and ${H^0}$ as the query, passing through the cross-attention module to compute the next hidden state ${H^1}$. To reduce the learning difficulty, we use residual connections so that ${H^1}$ represents information about changes in TC intensity. A feed-forward networks (FFN) module processes ${H^1}$ and adds it to the intensity ${I_{i+3}}$ of the previous time step to forecast the intensity ${I_{i+4}}$ of the next time step. This process repeats $j$ times to obtain forecasts of TC intensity at $j$ time steps. The described process is represented as follows:
\begin{align}
    ~{I_{i+N+j-1}}, ~{H^{j}} =~{AR}~{Decoder}(~{I_{i+N+j-2}}, ~{H^{j-1}})
    \label{eq:AR_Decoder}
\end{align}
where $j\in [1, 2, \cdots ,M]$ and ${N}$ represents the input length.

\section{Experiments}
To demonstrate the superiority and robustness of the MSCAR model, as well as the availability of the SETCD dataset, we conduct a series of experiments. These experiments include global short-term TC intensity forecasting, regional short-term TC intensity forecasting, real-time forecasting, and ablation experiments.

\subsection{Experimental Setup}

\paragraph{Datasets} Due to missing values in the water vapor and visible channels, early satellite imagery had limited available samples. To maximize the model's performance and learn from diverse TC morphologies, we focus on utilizing the infrared channel, which provides relatively more complete data, in our experiments. In contrast, all ERA5 variables are complete, so we employ all 69 dimensions of this environmental data. Based on~\cite{TC_Pred}, considering that the scale of TCs can reach up to 1000 km, we select satellite data and ERA5 data from the SETCD dataset within a diameter of approximately $10^\circ$ centered on the TC. The dimensions of a single example of a satellite image are (1, 140, 140), while the dimensions of ERA5 data are (69, 40, 40).

Deviation in TC intensity exists among different agencies~\cite{IBTrACS}. To minimize these discrepancies and include a larger number of samples for training, we adopt the the ``USA\_WIND" (knots) and ``USA\_PRES" (hPa) from the IBTrACS dataset as the ground truth for MSW and MSLP respectively. For convenient comparison, we have standardized the wind speed unit to meters per second (m$/$s), where 1 knot is equivalent to 0.5144444 m$/$s. Since the majority of the IBTrACS dataset provides optimal estimates of TC intensity at 6-hour intervals~\cite{IBTrACS}, we also adopt a 6-hour time resolution for sampling in the SETCD dataset. The data is partitioned as follows: the training set spans from 1980 to 2017, the validation set covers data from 2018, the test set incorporates data from 2019 to 2020, and the real-time performance evaluation test set includes data from 2021 to 2022.

\begin{table*}
    \resizebox{\textwidth}{!}{
    \centering
    \begin{tabularx}{\linewidth}{l
    >{\raggedleft\arraybackslash}X
    >{\raggedleft\arraybackslash}X
    >{\raggedleft\arraybackslash}X
    >{\raggedleft\arraybackslash}X
    >{\raggedleft\arraybackslash}X
    >{\raggedleft\arraybackslash}X
    >{\raggedleft\arraybackslash}X
    >{\raggedleft\arraybackslash}X
    }
        \toprule
        \multirow{2.5}{*}{Methods}    & \multicolumn{4}{c}{MSW (m$/$s)}     & \multicolumn{4}{c}{MSLP (hPa)} \\
         \cmidrule{2-9}
              &  6h   & 12h   & 18h   & 24h   &  6h   & 12h   & 18h   & 24h\\
        \midrule
        ECMWF & 7.34  & 7.29  & 7.43  & 7.48  & 6.73  & 6.85  & 7.27  & 7.65 \\
        NCEP  & 5.09  & 5.33  & 5.58  & 5.95  & 4.87  & 5.65  & 6.40   & 6.98 \\
        LSTM  & 1.79  & 3.22  & 4.56  & 5.79  & 2.41  & 4.19  & 5.86  & 7.39 \\
        Transformer & 1.91  & 3.34  & 4.66  & 5.85  & 2.59  & 4.36  & 6.04  & 7.53 \\
        ConvGRU & 1.79  & 2.83  & 3.77  & 4.58  & 2.52  & 3.78  & 5.01  & 6.06 \\
        MSCAR & \textbf{1.64}  & \textbf{2.72}  & \textbf{3.56}  & \textbf{4.30}   & \textbf{2.28}  & \textbf{3.60}   & \textbf{4.66}  & \textbf{5.62} \\

        \bottomrule
    \end{tabularx}}
    \caption{Comparison of the average absolute error in global short-term TC intensity forecast between 2019 and 2020.}
    \label{tab:Global}
\end{table*}

\paragraph{Implementation Details} The MSCAR model is trained in the PyTorch framework using 8 Nvidia A100 GPUs. The model is optimized using AdamW with an initial learning rate of 0.0001. The training process consists of 50 epochs with a batch size of 8, employing the MAE loss function. An L1 regularization term is incorporated with a regularization coefficient of 0.00001 to mitigate overfitting. Additionally, the Exponential Moving Average (EMA) technique is used as a regularization mechanism. As depicted in Figure \ref{fig:MSCAR}, the data from the past 24 hours (N=4) is utilized to forecast the forthcoming TC intensity values.  While MSCAR allows for experimenting with forecasts at arbitrary time steps, our focus is on short-term intensity forecasting for TCs over the next 24 hours (M=4). For more detailed information, please refer to the Appendix section of our study.

\subsection{Global Short-Term TC Intensity Forecasting}
To validate MSCAR's superiority, performance testing and comparison are conducted based on the average absolute intensity error on the global TCs dataset from 2019 to 2020, as presented in Table \ref{tab:Global}. A comparative analysis uses global models ECMWF-IFS and NCEP-GFS, renowned official forecasting agencies. The forecast data for these models are obtained from ~\cite{TIGGE}. Furthermore, the methods of LSTM~\cite{LSTM}, Transformer~\cite{transformer_track_intensity_forecasting}, and ConvGRU~\cite{TC_Pred} are reproduced. For specific details regarding modifications and adjustments, please refer to the Appendix.

Based on the experimental results, it can be observed that ECMWF-IFS and NCEP-GFS exhibit subpar performance. However, due to their physical constraints, these models demonstrate remarkably slow growth in forecast errors over time, which makes them more suitable for medium and long-term forecasting. LSTM and Transformer models have shown some advancements in short-term forecast performance, yet their sole reliance on time series information hinders their ability to account for spatial variations. As the forecasting horizon increases, their performance deteriorates significantly. By incorporating spatial information, ConvGRU demonstrates further performance improvements. However, it lacks sensitivity to TC scale variations and is challenging to extend to longer forecast time steps.

MSCAR achieves the best short-term forecasting performance, with MSW showing significant improvements compared to previous models at 24-hour intervals, with enhancements of 42.51\%, 27.73\%, 25.73\%, 26.50\%, and 6.11\%. Compared to ConvGRU, MSCAR demonstrates an overall performance improvement of 3.89\% to 9.52\%. Moreover, MSCAR is more flexible as it allows for forecasting at arbitrary time steps.

\begin{table}
    \centering
    
    \begin{tabular}{llrrrr}
        \toprule
        \multirow{2.5}{*}{Methods} & \multirow{2.5}{*}{Basin} & \multicolumn{4}{c}{MSW (m$/$s)} \\

        \cmidrule{3-6} & &  6h   & 12h   & 18h   & 24h \\
        \midrule
        SHIFOR5 & \multirow{6}{*}{AL} & {---} & 3.10   & {---} & 5.35 \\
        OFCL  &       & {---} & 2.83  & {---} & 4.30 \\
        LSTM  &       & 1.63  & 3.12  & 4.49  & 5.65 \\
        Transformer &       & 1.79  & 3.25  & 4.60   & 5.67 \\
        ConvGRU &       & 1.68  & 2.72  & 3.65  & 4.38 \\
        MSCAR &       & \textbf{1.54}  & \textbf{2.68}  & \textbf{3.50}   & \textbf{4.09} \\
        \midrule
        SHIFOR5 & \multirow{6}{*}{EP} & {---} & 3.03  & {---} & 5.56 \\
        OFCL  &       & {---} & 2.64  & {---} & 4.09 \\
        LSTM  &       & 1.53  & 2.83  & 4.08  & 5.23 \\
        Transformer &       & 1.63  & 2.89  & 4.07  & 5.20 \\
        ConvGRU &       & 1.47  & 2.35  & 3.14  & 3.80 \\
        MSCAR &       & \textbf{1.40}   & \textbf{2.31}  & \textbf{3.05}  & \textbf{3.61} \\
        \midrule
        JTWC real-time & \multirow{8}{*}{WP} & {---} & {---} & {---} & 4.60 \\
        CMA real-time &       & {---} & {---} & {---} & 4.80 \\
        LSTM  &       & 1.74  & 3.19  & 4.56  & 5.86 \\
        Transformer &       & 1.89  & 3.36  & 4.75  & 6.04 \\
        ConvGRU &       & 1.89  & 2.97  & 3.98  & 4.79 \\
        MMSTN &       & 2.24  & 4.13  & 6.14  & 8.18 \\
        MGTCF &       & 2.15  & 3.68  & 5.11  & 6.21 \\
        MSCAR &       & \textbf{1.58}  & \textbf{2.62}  & \textbf{3.49}  & \textbf{4.29} \\
        \bottomrule 
    \end{tabular}
    \caption{Comparative analysis of average absolute error in short-term TC intensity forecast between 2019 and 2020 in the AL (Atlantic) and EP (Eastern Pacific) basins, alongside average absolute error for 2019 short-term TC intensity forecast in the WP (Western Pacific) basin. ``---'' denotes missing data.}
    \label{tab:Regional}
\end{table}
\subsection{Regional Short-Term TC Intensity Forecasting}
With a focus on minimizing disasters, regions worldwide often prioritize the study of TC in their neighboring marine areas. As a result, we have selected three prominent marine regions, including the Atlantic (AL), Eastern North Pacific (EP), and Western North Pacific (WP). The results are presented in Table~\ref{tab:Regional}. In the AL basin and EP basin, the statistical baseline model SHIFOR5 from the NHC and the official forecast OFCL provided by the NHC are selected for comparison. The performance comparison is conducted for the years 2019 and 2020, using data sourced from the official website of the NHC. For the WP basin, the forecast results from JTWC real-time and CMA real-time~\cite{chen2019verification}, along with the deep learning models MMSTN~\cite{mmstn} and MGTCF~\cite{mgtcf}, are chosen. Due to the unique nature of the input data for MMSTN and MGTCF models, we specifically compare their performance in the WP basin for 2019. It is important to note that, to ensure fairness in the deterministic forecasting experiments and error calculations for MMSTN and MGTCF, the average results from multiple prediction samples are used instead of selecting a single sample with the smallest error compared to the ground truth, as mentioned in the original paper. This approach is adopted because, in practical forecasting tasks, it is impossible to know which sample would have the smallest error.

Due to official TC intensity forecasts typically using MSW, Table~\ref{tab:Regional} primarily showcases the performance of MSW forecasts. In various basins, there are notable differences in the intensity and quantity of TCs due to factors such as variances in ocean temperatures, atmospheric circulation patterns, and conditions for TC formation. The ConvGRU model maintains good performance in the AL and EP basins, but it noticeably lags behind the MSCAR model in the more complex WP basin, where TC activity is more prominent. The main reason for this difference lies in the multi-scale information extraction employed by the MSCAR model, which is more sensitive to TCs of different scales compared to the ConvGRU model's sole reliance on convolution operations. The results of MMSTN and MGTCF are not satisfactory, indicating that while they may serve as ensemble forecasts to provide certain early warning capabilities, they are unsuitable for deterministic forecasting tasks. The results of the experiments demonstrate that MSCAR consistently achieves the best performance across all three basins and is the only deep-learning model that surpasses all the aforementioned official forecasts. Specifically, MSCAR outperforms OFCL by 4.88\% in the AL basin, 11.74\% in the EP basin, and surpasses JTWC by 6.74\% and CMA by 10.63\% in the WP basin, all within a 24-hour prediction interval. 

\subsection{Real-time Forecasting}

It is important to note that we are not comparing the models in real-time conditions since real-time access to ERA5 data is not available. To validate the practical value of MSCAR, we replace the input ERA5 data with analysis data of the same variables, which can be considered as near real-time data. 

The results are tested before and after the data replacement for 2021 and 2022, as shown in Table~\ref{tab:Real_time}.  There is little noticeable change in performance, reaffirming MSCAR's ability to maintain superior performance in real-time forecasting while also highlighting its robustness. %Interestingly, a fascinating phenomenon is observed where the TC intensity forecast errors produced by MSCAR remain consistently stable across different years, with performance for each year remarkably similar, indicating the model's consistent and reliable performance. For more detailed results, please refer to the Appendix.

\begin{table}
    \centering
    \begin{tabular}{lrrrr}
        \toprule
        \multirow{2.5}{*}{Methods}  & \multicolumn{4}{c}{MSW (m$/$s)} \\
         \cmidrule{2-5}
                               &  6h   & 12h   & 18h   & 24h \\
        \midrule
        MSCAR                  & \textbf{1.72}  & 2.74           & \textbf{3.57}  & \textbf{4.29}  \\
        MSCAR(real-time)       & \textbf{1.72}  & \textbf{2.73}  & 3.59  & 4.31  \\
        \bottomrule 
    \end{tabular}
    \caption{Real-time forecast performance testing of MSCAR for 2021 and 2022.}
    \label{tab:Real_time}
\end{table}

\subsection{Ablation Experiments}
From the previous experiments, it can be observed that the MSCAR model successfully extracts valuable information from satellite data and ERA5 data, resulting in improved performance.
In this study, an ablation experiment is conducted to explore the impact of different inputs. Specifically, one of the following inputs is removed: satellite data, ERA5, or historical TC intensity sequences, and then the MSCAR model is retrained. Performance is evaluated using global TC data from 2019 and 2020. The results are presented in Table ~\ref{tab:Ablation_datasets}. In addition to the data ablation experiments, model structure ablation experiments are also conducted to understand the contribution of different components. The results are presented in Table~\ref{tab:Ablation_model}.

\paragraph{The Impact of Different Inputs} 
The high-resolution spatial features captured by satellite imagery benefit TC intensity forecasting efforts seeking improved accuracy. ERA5 plays a crucial role in our MSCAR model, and its integration leads to a significant performance improvement ranging from 9.39\% to 25.09\%. This highlights the complexity of TCs as a highly intricate atmospheric system, with various degrees of coupling among multiple atmospheric variables. The historical TC intensity sequence also plays a crucial role, particularly for one-step and two-step forecasts, resulting in improvements of 54.95\% and 33.33\% respectively. Generally, in one-step forecasts, the TC intensity does not change dramatically. By leveraging the knowledge of the current step's value, significant error reduction can be achieved through the use of residual connections. Therefore, the historical TC intensity sequence is indispensable for the MSCAR model.

\begin{comment}
\paragraph{The Role of Satellite Data} When we remove satellite data, the performance does not show a significant decline, which is actually in line with expectations. This is because there is a certain overlap between the cloud features present in satellite data and the feature information in ERA5 data. Additionally, we only use one channel of satellite data, whereas ERA5 contains 69 variables, making it easier for the extracted information to be overshadowed. However, satellite data provides higher-resolution feature information, which can contribute to further accuracy improvements. If we can utilize more channels of satellite data, it is possible that satellite data could play a more significant role in enhancing the model's performance. 
\end{comment}

\paragraph{The Impact of Different Model Structure} 
The ablation experiments aim to analyze the roles of FPN, CC Attention, and AR Decoder in the MSCAR model. 

After replacing FPN with U-Net~\cite{U-net}, the overall forecast performance decreases, with a 4.19\% drop for the 24-hour lead time. Both U-Net and FPN are modules that fuse multi-scale features, but U-Net only utilizes the final fused feature map, while FPN retains more feature maps at different scales, making it easier for FPN to capture crucial small-scale information, such as the characteristics of the eye of a TC.

CC Attention is a key component of MSCAR. When we replace the causal cross-attention computation with self-attention computation, the overall performance significantly declines, with a 22.09\% decrease for the 24-hour lead time. This confirms the importance of prior temporal causal relationships. Directly using self-attention not only increases computational complexity but also introduces redundant information, hindering effective feature extraction and leading to potential overfitting.

AR Decoder is designed to ensure performance while achieving arbitrary-step forecasts, enhancing the model's flexibility. Here, we do not modify the network structure of the AR Decoder but only change the output from one time step to four-time steps, eliminating the autoregressive process and transforming it into a MIMO structure. Experimental results demonstrate that AR Decoder achieves our goal and exhibits improved performance, with a 2.93\% increase in performance for the 24-hour lead time.

\begin{table}
    \centering
    \begin{tabular}{lllrrrr}
        \toprule
        \multirow{2.5}{*}{${S}$}  & \multirow{2.5}{*}{${E}$}  & \multirow{2.5}{*}{${I}$}  & \multicolumn{4}{c}{MSW (m$/$s)} \\
         \cmidrule{4-7}
                &        &       &  6h   & 12h   & 18h   & 24h \\
        \midrule
         &     $\checkmark$   &    $\checkmark$     & 1.64  & 2.75  & 3.60  & 4.38  \\
        $\checkmark$  &      &     $\checkmark$     & 1.81  & 3.19  & 4.51  & 5.74  \\
        $\checkmark$  &  $\checkmark$   &       & 3.64  & 4.08  & 4.57  & 5.11  \\
        $\checkmark$  &   $\checkmark$  &  $\checkmark$    & \textbf{1.64}  & \textbf{2.72}  & \textbf{3.56}  & \textbf{4.30}  \\
        \bottomrule 
    \end{tabular}
    \caption{Ablation experiments: the impact of satellite data (${S}$), ERA5 data (${E}$), and historical TC intensity sequences (${I}$).}
    \label{tab:Ablation_datasets}
\end{table}

\begin{table}
    \centering
    \begin{tabular}{lrrrr}
        \toprule
        \multirow{2.5}{*}{Methods}   & \multicolumn{4}{c}{MSW (m$/$s)} \\
        \cmidrule{2-5}
                           &  6h   & 12h   & 18h   & 24h \\
        \midrule
        U-Net+CC Atten+AR        & 1.65  & 2.76  & 3.67  & 4.48  \\
        FPN+Self Atten+AR       & 1.79 	&3.10	&4.24	&5.25  \\
        FPN+CC Atten+MIMO       & 1.69 	&2.77	&3.64	&4.43 \\
        FPN+CC Atten+AR         & \textbf{1.64}  & \textbf{2.72}  & \textbf{3.56}  & \textbf{4.30}  \\
        \bottomrule 
    \end{tabular}
    \caption{Ablation experiments: the impact of FPN, CC Attention(CC Atten), and AR Decoder(AR).}
    \label{tab:Ablation_model}
\end{table}

\section{Conclusion}
In this paper, we present MSCAR, a global TC intensity forecasting model that incorporates prior causal relationships, extracts multi-scale features from a large-scale multimodal dataset, and employs autoregressive modeling in latent space. Additionally, we introduce SETCD, a novel dataset encompassing a wide range of spatial variables for global TC analysis. MSCAR demonstrates state-of-the-art performance both globally and within regional basins, exhibiting significant robustness and practical value. The effectiveness of SETCD has also been confirmed, and we anticipate its valuable contributions to future TC research.

\section*{Acknowledgements}
We express our utmost gratitude to ECMWF and NOAA for providing us with valuable datasets, which have been instrumental in our work. We would also like to extend our sincere appreciation to the Shanghai AI Laboratory for their technical guidance and the provision of computing resources. Without their invaluable support, our research would not have been possible.

This work is funded by the National Natural Science Foundation of China (U19B2044), supported by Shanghai Science and Technology Commission Project (23DZ1204704).

\appendix

\bibliographystyle{named}
\bibliography{ijcai24}

\newpage

\section{Appendix}

\subsection{SETCD}

Each sample in the SETCD dataset consists of two npy files: GRIDSAT\_data.npy and ERA5\_data.npy. Moreover, to address the potential issue of duplicate occurrences of the same TC name, SETCD utilizes a unique storm identifier (SID) assigned by the IBTrACS algorithm as the naming convention for storage. The variables stored in these files and their respective order are presented in Table~\ref{tab:SETCD_Variable}. 

\begin{table}[htbp]
    {
    \centering
    \begin{tabularx}{\linewidth}{ll}
        \toprule
        Data    & Sample Variable Composition   \\
        
        \midrule
        \multirow{11}{*}{ERA5} & [u10, v10, t2m, msl,\\
                                           &    z50, z100, z150, z200, z250, z300, z400,\\
                                           &    z500, z600, z700, z850, z925, z1000,\\
                                           &    q50, q100, q150, q200, q250, q300, q400, \\
                                           &    q500, q600, q700, q850, q925, q1000,\\
                                           &    u50, u100, u150, u200, u250, u300, u400, \\
                                           &    u500, u600, u700, u850, u925, u1000,\\
                                           &    v50, v100, v150, v200, v250, v300, v400, \\
                                           &    v500, v600, v700, v850, v925, v1000,\\
                                           &    t50, t100, t150, t200, t250, t300, t400, \\
                                           &    t500, t600, t700, t850, t925, t1000]\\
        GRIDSAT  & [irwin\_cdr, irwvp, vschn]\\
        \bottomrule
    \end{tabularx}}
    \caption{Composition and order of variables in individual samples of the SETCD dataset.}
    \label{tab:SETCD_Variable}
\end{table}

Here is a brief introduction to GridSat-B1 and ERA5:
\paragraph{GridSat-B1} The Geostationary IR Channel Brightness Temperature (BT)- GridSat-B1 Climate Data Record (CDR) ~\cite{gridsat1,gridsat2}, with a temporal resolution of 3 hours and a spatial resolution of 0.07$^\circ$$\times$0.07$^\circ$, encompasses global (70N to 70S) infrared window brightness temperatures from geostationary Infrared (IR) satellites, spanning from 1980 to the present day. It comprises the following three channels:
\begin{itemize}
    \item CDR-quality infrared window channel (near 11 $\upmu$m).
    \item Infrared water vapor channel (near 6.7 $\upmu$m).
    \item Visible channel (near 0.6 $\upmu$m).
\end{itemize}

\paragraph{ERA5} The fifth-generation ECMWF atmospheric reanalysis, represents a comprehensive global climate dataset from January 1940 to the present. With a spatial resolution of 30km, ERA5 offers hourly estimates of numerous atmospheric, land, and oceanic climate variables. It utilizes 137 vertical levels, covering the Earth's atmosphere from the surface up to 80km in height, ensuring a high-resolution representation of atmospheric conditions. 

\begin{figure}
    \centering
    \includegraphics[width=0.45\textwidth]{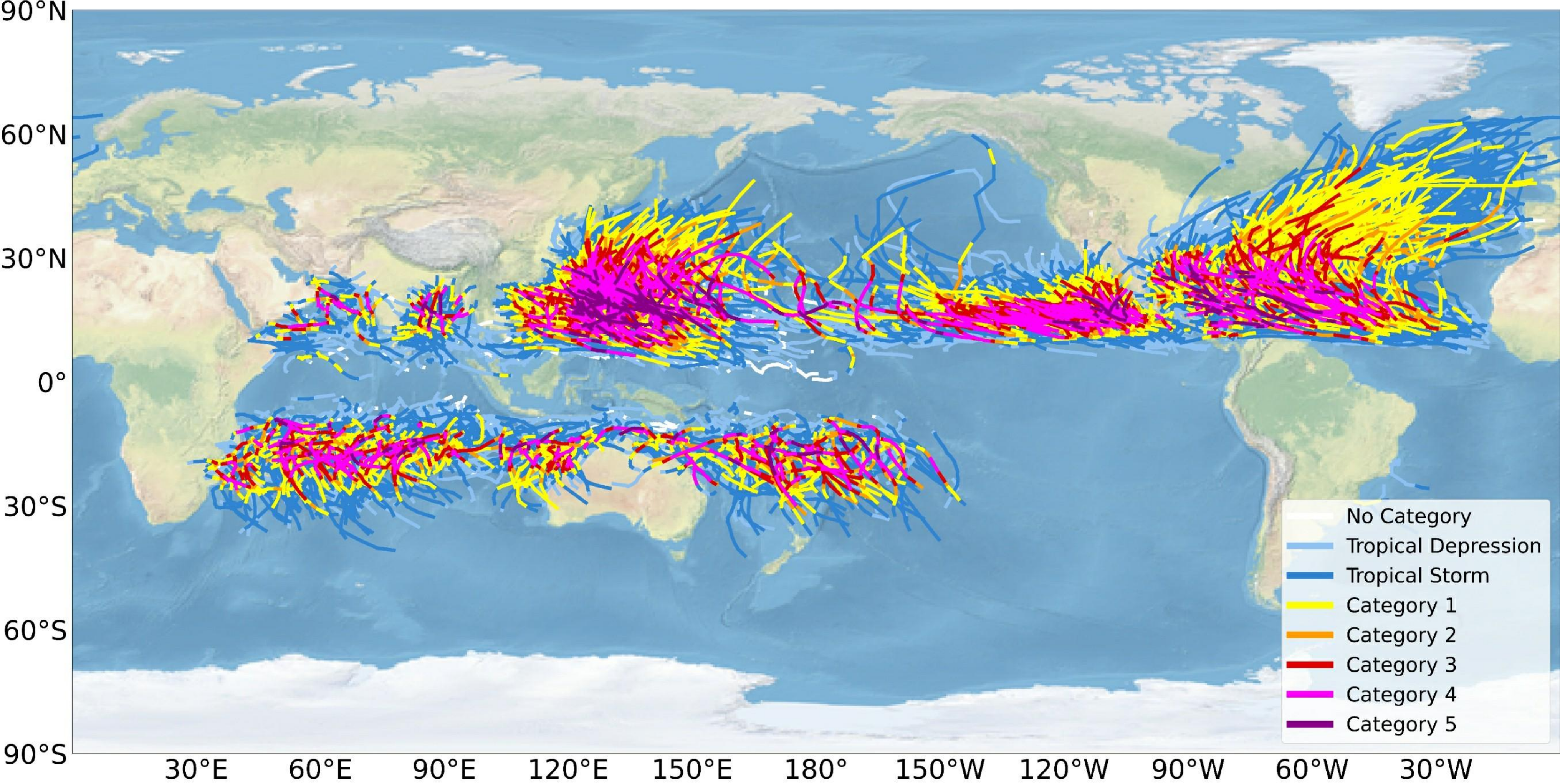}
    \caption{The global TC colored by their Saffir-Simpson Hurricane Wind Scale (SSHWS) intensity categories from 1980 to 2022.}
    \label{fig:all_tc}
\end{figure}

\begin{table}
    \centering
    \begin{tabular}{lr}
        \toprule
        Category  & MSW\\
        \midrule
        No Category             & $\le$ 20 knots\\
        Tropical Depression     & 21-33 knots\\
        Tropical Storm          & 34-63 knots\\
        Category 1              & 64-82 knots\\
        Category 2              & 83-95 knots \\
        Category 3              & 96-112 knots \\
        Category 4              & 113-136 knots \\
        Category 5              & $\ge$ 137 knots \\
        \bottomrule 
    \end{tabular}
    \caption{Saffir-Simpson Hurricane Wind Scale (SSHWS).}
    \label{tab: SSHWS}
\end{table}

\begin{figure*}
    \centering
    \includegraphics[width=0.98\textwidth]{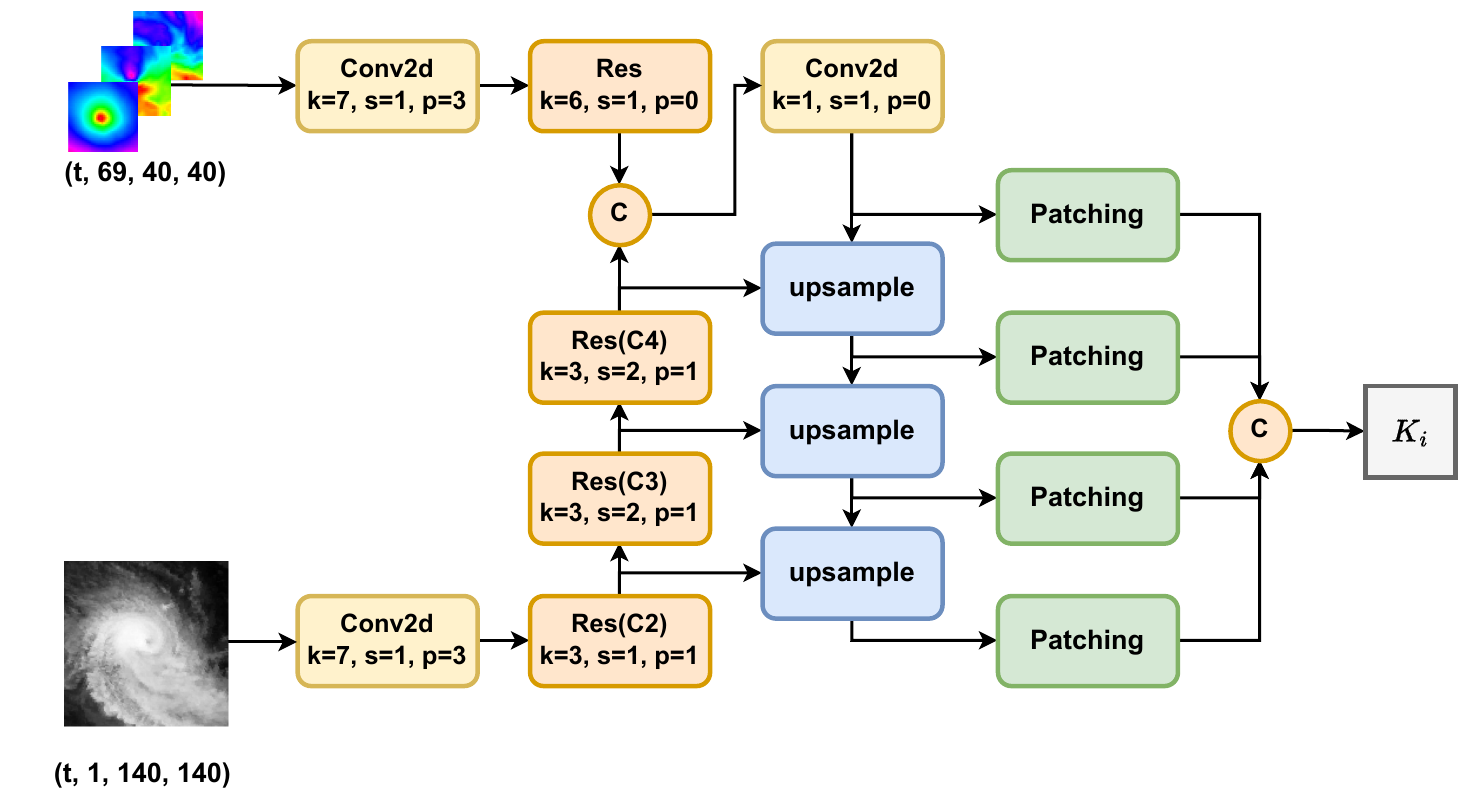}
    \caption{The FPN structure framework diagram we used in the experiment. ``Res'' represents the ResNets module~\protect\cite{resnet}, ``k'' denotes the kernel size, ``s'' indicates the stride, and ``p'' represents the padding.}
    \label{fig:FPN}
\end{figure*}

We conduct a statistical analysis of the NaN value proportion for each sample in the satellite data with a diameter of $10^\circ$ during the experimental process. The infrared and water vapor channels have a completeness rate of over 90\%, indicating a high level of data availability. However, the visible channel is approximately 35\% complete, suggesting some missing values. To address these incomplete portions, all missing values have been uniformly treated and replaced with NaN (not a number) values during data processing. We provide a statistical file, where data with NaN value proportions below 1\% are typically used by default for the experiment.

Figure~\ref{fig:all_tc} presents all TCs and their intensity categories involved in this study. The intensities are classified according to the Saffir-Simpson Hurricane Wind Scale (SSHWS) as shown in Table~\ref{tab: SSHWS}, and the data is sourced from IBTrACS.

\subsection{Some Details of the MSCAR Model}

\paragraph{FPN}Figure~\ref{fig:FPN} illustrates the detailed structure and important parameters of the FPN in our experiment. It differs from the original FPN in two main aspects. Firstly, the satellite images and ERA5 data are downsampled to the same scale before fusing the information. Secondly, the feature maps of different scales are segmented into fixed-size patches, linearly embedded, and then concatenated to obtain the spatial information token ${K}$.

\begin{figure}[htbp]
    \centering
    \includegraphics[width=0.45\textwidth]{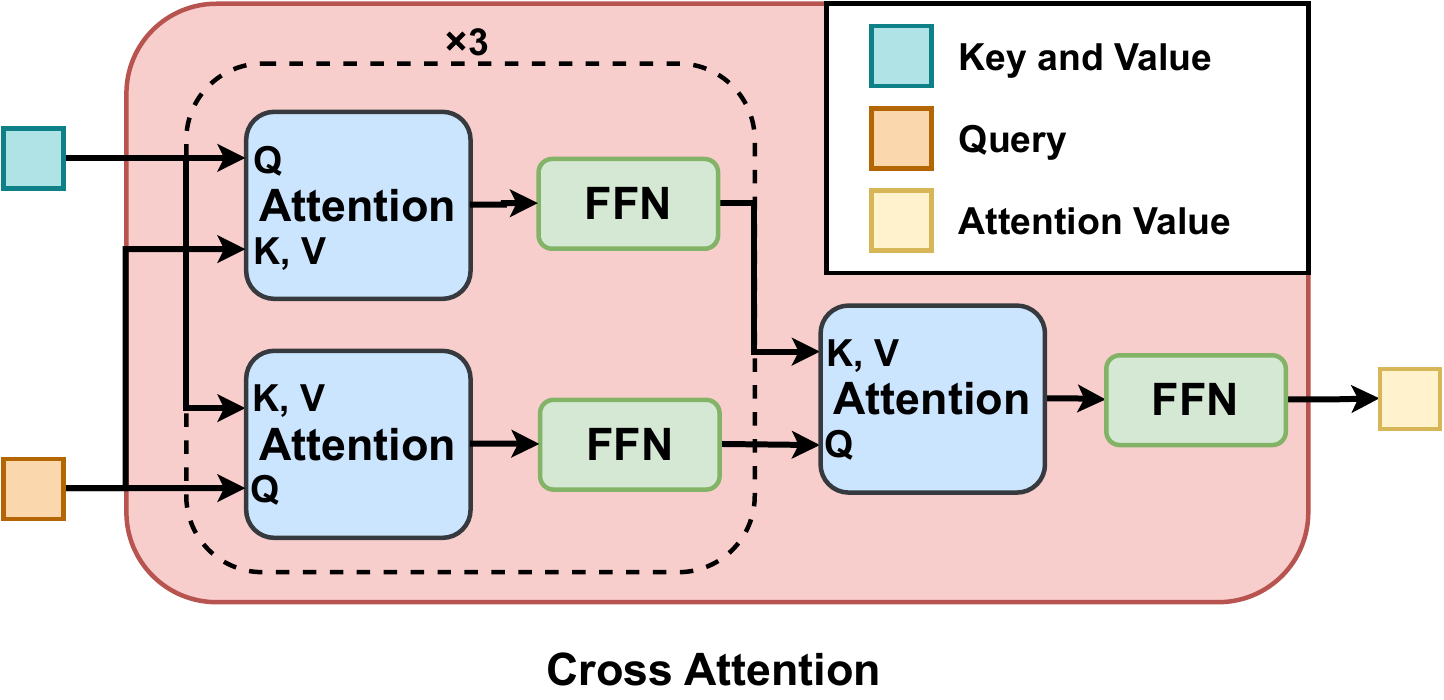}
    \caption{The figure illustrates the computation process of the Cross Attention module, where ``Attention'' represents the multi-head attention calculation module, and ``FFN'' stands for the feed-forward neural network.}
    \label{fig:Cross_att}
\end{figure}

\paragraph{Cross Attention Module} In the MSCAR model, the Cross Attention Module involves multiple iterations of cross-attention calculations, as depicted in Figure~\ref{fig:Cross_att}. It is not a simple one-time cross-attention computation. The Attention Module indeed utilizes multi-head attention calculations.

\subsection{Model Reproduction and Training Details}
In this study, the LSTM model uses the original LSTM architecture. The Transformer model replicates the approach described in~\cite{transformer_track_intensity_forecasting}, with the input set to 4. The ConvGRU model is based on the TC-Pred framework proposed in~\cite{TC_Pred}, incorporating the ConvGRU architecture from~\cite{convgru} as its encoder-decoder framework. During the replication process, we found that the feature enhancement module of TC-Pred does not improve the performance of the SETCD dataset. As a result, the ConvGRU model used in the experiment excludes this module. The original papers of MMSTN~\cite{mmstn} and MGTCF~\cite{mgtcf} provide the code and corresponding data, allowing us to directly utilize the original code in our experiments.

The experimental data is normalized and our code provides the mean and variance values for all variables. It is important to note that in the paper, ``USA WIND'' (knots) and ``USA PRES'' are chosen as the Ground Truth for Maximum Sustained Wind (MSW) and Mean Sea Level Pressure (MSLP) respectively. If data from other meteorological agencies is selected as the Ground Truth, re-calculating the mean and variance is necessary due to differences in definitions and techniques. This can potentially have a significant impact on the results.

\begin{table*}[htbp]
    \resizebox{\textwidth}{!}{
    \centering
    \begin{tabularx}{\linewidth}{l
    >{\raggedleft\arraybackslash}X
    >{\raggedleft\arraybackslash}X
    >{\raggedleft\arraybackslash}X
    >{\raggedleft\arraybackslash}X
    >{\raggedleft\arraybackslash}X
    >{\raggedleft\arraybackslash}X
    >{\raggedleft\arraybackslash}X
    >{\raggedleft\arraybackslash}X}
        \toprule
        \multirow{2.5}{*}{Year}    & \multicolumn{4}{c}{MSW (m$/$s)}     & \multicolumn{4}{c}{MSLP (hPa)} \\
         \cmidrule{2-9}
                &  6h   & 12h   & 18h   & 24h   &  6h   & 12h   & 18h   & 24h\\
        \midrule
        2018    &1.54  &2.61  &3.48 & 4.20          &2.24 & 3.71 & 4.93 & 5.90 \\
        2019    &1.62 & 2.71 & 3.61 & 4.39          &2.36 & 3.73 & 4.87 & 5.93\\
        2020    &1.66 & 2.74 & 3.50 & 4.18          &2.18 & 3.44 & 4.40 & 5.24 \\
        2021    &1.74 & 2.74 & 3.59 & 4.30          &2.41 & 3.61 & 4.73 & 5.62 \\
        2022    &1.70 & 2.72 & 3.52 & 4.26          &2.34 & 3.77 & 4.94 & 6.10 \\

        \bottomrule
    \end{tabularx}}
    \caption{Historical global TC intensity short-term forecast errors for MSCAR model from 2018 to 2022.}
    \label{tab:yearS_error}
\end{table*}

\subsection{Additional Experimental Results}
\subsubsection{Comparative Analysis of Medium to Long-Term Forecast Performance}
MSCAR is an autoregressive forecast model capable of achieving predictions at any time step, we conducted experiments for medium and long-term global TC intensity forecasting in 2018.  A comparison is made between the results obtained by MSCAR and those presented in~\cite{cmp_pangu}, as displayed in Table~\ref{tab: Medium to Long-Term}. The experimental results demonstrate that MSCAR exhibits the best performance for a forecast horizon of approximately 48 hours. 
%As the forecast horizon extends, the prediction error increases, and MSCAR's advantage diminishes. The main reason for the poor long-term forecasting performance of MSCAR is its excessive autoregressive iterations, leading to cumulative errors. In contrast, IFS incorporates physical constraints, ensuring that its predictions adhere to physical laws, resulting in a slower increase in prediction error with longer forecast horizons. The Pangu model, seemingly learning some form of constraint from a large set of atmospheric variables, also shows a slower growth in prediction error. However, due to the tendency of ERA5 data to underestimate TC intensity, Pangu's atmospheric field predictions naturally underestimate TC intensity as well, leading to significant predictive biases in its initial forecasts.

\begin{table}[H]
    \centering
    \begin{tabular}{lrr}
        \toprule
        \multirow{2.5}{*}{Methods}  & \multicolumn{2}{c}{MSLP (hPa)} \\
         \cmidrule{2-3}
                  &  24h   & 48h     \\
        \midrule
        
        IFS       & 9.42  & 11.00   \\
        Pangu~\cite{pangu}     & 17.47 & 18.11  \\
        MSCAR     & \textbf{5.90}  & \textbf{10.33}   \\
        \bottomrule 
    \end{tabular}
    \caption{Comparison of forecast error performance for global TC intensity in 2018. It should be noted that~\protect\cite{cmp_pangu} only evaluated model performance for MSLP forecasts. Therefore, this work only presents comparative results for MSLP prediction skills.}
    \label{tab: Medium to Long-Term}
\end{table}

For regional performance comparison, we also compare the performance of MSCAR with five other global operational forecasting models in the 2019 Western Pacific (WP) basin for medium to long-term forecasts. The five global operational forecasting models are respectively: Integrated Forecasting System of ECMWF (ECMWF-IFS), Global Spectral Model of JMA (JMA-GSM ), Global Forecast System of NCEP (NCEP-GFS), Global Data Assimilation and Prediction System of KMA (KMA-GDAPS), UKMO-MetUM Unified Model system of UKMO (UKMO-MetUM). Similar to the results in Table~\ref{tab: Medium to Long-Term}, we obtain comparable findings, where MSCAR achieves state-of-the-art performance within a 48-hour forecast horizon, as shown in Table~\ref{tab: Medium to Long-Term regional}.

\begin{table}[htbp]
    \centering
    \begin{tabular}{lrr}
        \toprule
        \multirow{2.5}{*}{Methods}  & \multicolumn{2}{c}{MSW (m$/$s)} \\
         \cmidrule{2-3}
                    & 24h   & 48h   \\
        \midrule
        
        ECMWF-IFS   &7.30	&9.10	  \\
        JMA-GSM     &7.60	&11.50	 \\
        NCEP-GFS    &7.20	&9.10	 \\
        KMA-GDAPS   &5.90	&8.50	 \\
        UKMO-MetUM  &10.50	&11.90		\\
        MSCAR       &\textbf{4.29}   &\textbf{8.45}    \\
        \bottomrule 
    \end{tabular}
    \caption{Comparison of forecast error performance for TC intensity in the Western Pacific basin in 2019.}
    \label{tab: Medium to Long-Term regional}
\end{table}

\subsubsection{Case Study}
\begin{figure*}[!htb]
    \centering
    \subfigure[LANE (SID 2018226N11245)]{\includegraphics[width=0.8\textwidth]{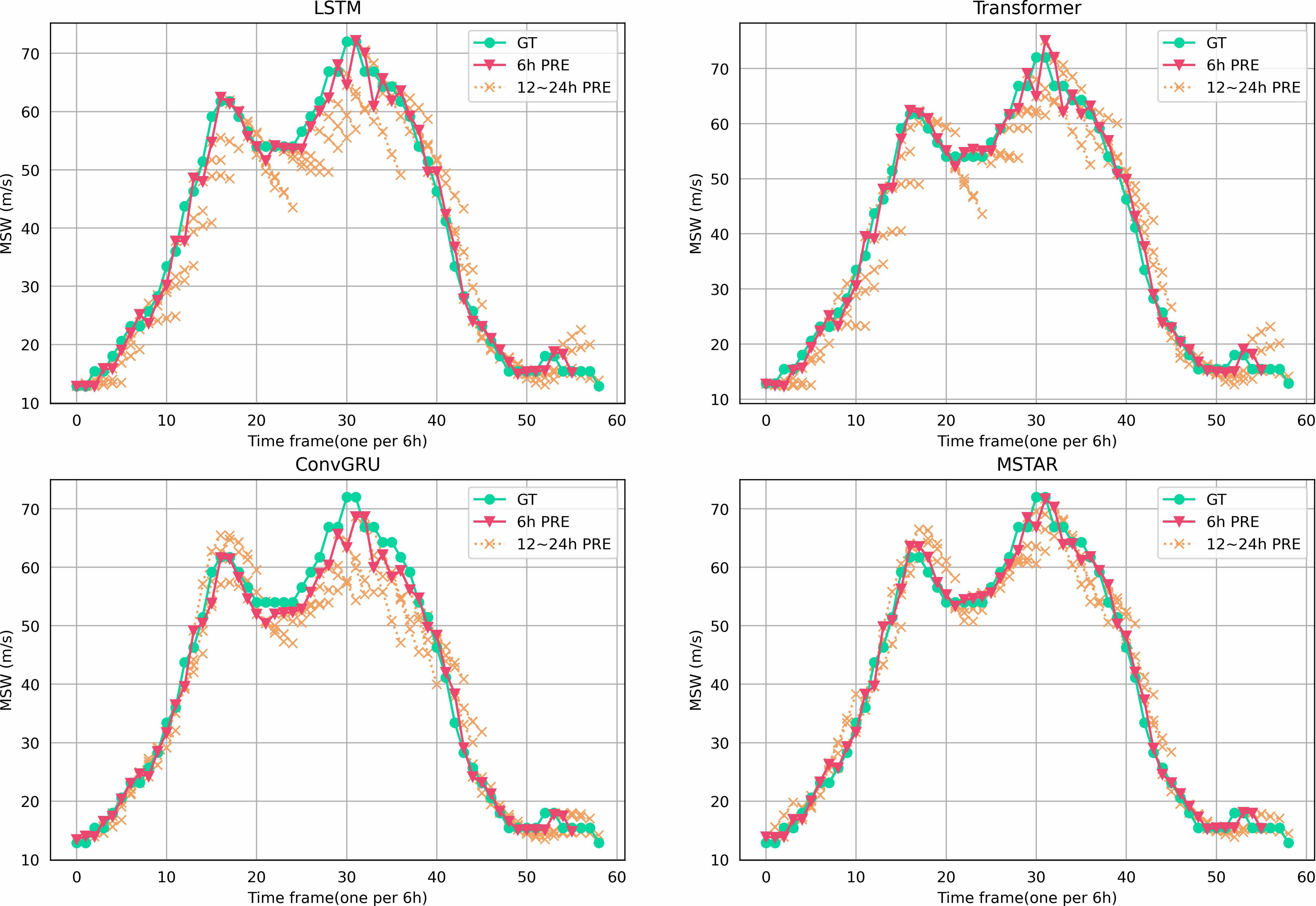} 
    \label{fig:2018226N11245}
    }
    \subfigure[HAISHEN (SID 2020244N25146)]{\includegraphics[width=0.8\textwidth]{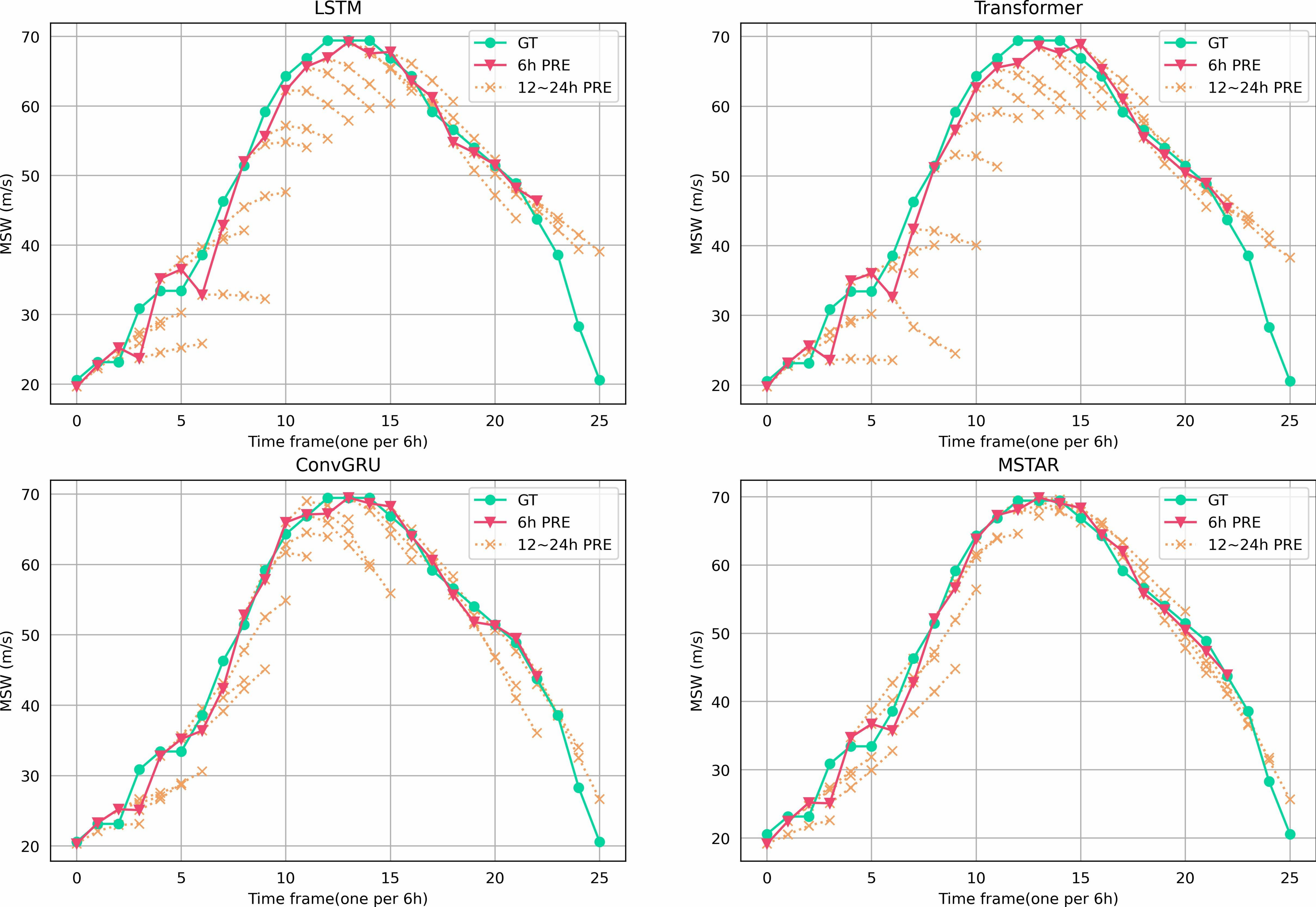}
    \label{fig:2020244N25146}
    }
    \caption{Visual comparison of TC intensity forecasts. ``GT'' represents the ground truth of TC intensity. ``6h PRE'' signifies the 6-hour forecast, while ``12$\sim$24h PRE'' denotes the forecast results for 12 hours, 18 hours, and 24 hours.}
\end{figure*}

In this section, we qualitatively analyze the performance differences among various models and present their forecasting results for TCs LANE (SID 2018226N11245) and HAISHEN (SID 2020244N25146), as illustrated in Figure~\ref{fig:2018226N11245} and ~\ref{fig:2020244N25146}.

LANE exhibits complex variations throughout its life cycle. During the initial rapid growth phase, both LSTM and Transform models perform poorly, showing overall weaker forecasts in the 12 to 24-hour range. In contrast, ConvGRU and MSCAR, which incorporate additional spatial information, effectively learn the growth trend of the TC and produce highly accurate forecasts from 6 to 24 hours. However, in the mid-term stage, when the TC reaches its first peak intensity, starts to decay, stabilizes for a period, and then undergoes a sudden intensification, ConvGRU performs poorly in predicting these complex multiple transition scenarios. Its one-step forecasts exhibit significant biases, and the 12 to 24-hour forecasts completely deviate from the actual trend. On the other hand, MSCAR successfully learns the dynamics of these intricate changes from the spatial information, resulting in much smaller overall trend errors.

HAISHEN has a relatively simple lifecycle. However, LSTM and Transform models still struggle to predict the growth of the TC and tend to underestimate its intensity. ConvGRU performs on par with MSCAR in terms of one-step predictions, but it faces challenges in accurately identifying the peak intensity position of the TC in the 12 to 24-hour forecasts. Once again, MSCAR demonstrates outstanding forecasting results, with its 6 to 24-hour forecasts providing more precise estimations of the TC's intensity peaks. This qualitative analysis highlights how MSCAR effectively learns the dynamic variations of the TC from historical information.

\subsubsection{Historical TC Intensity Forecast Performance of MSCAR}
Table~\ref{tab:yearS_error} illustrates the forecast error results of short-term intensity forecasts by MSCAR from 2018 to 2022. It can be observed that the forecast errors of MSCAR are relatively similar each year, indicating good stability in its performance.

\end{document}